\documentclass[paper,twocolumn,showpacs,showkeys,amsmath,amssymb]{revtex4}

\usepackage{graphicx}
\usepackage{setspace}
\usepackage{amsfonts}
\usepackage{amssymb,color}

\usepackage{graphicx,color}
\usepackage{ifpdf}
\usepackage[latin1]{inputenc}

\begin{document}

\title{Quenched disorder in the contact process on
  bipartite sublattices}

\author{M. N. Gonzaga$^{1}$, C. E. Fiore$^2$ and M. M. de Oliveira$^{1}$}
\address{
$^1$Departamento de F\'{\i}sica e Matem\'atica,
CAP, Universidade Federal de S\~ao Jo\~ao del Rei,
Ouro Branco-MG, 36420-000 Brazil, \\
$^2$ Instituto de F\'isica, Universidade de S\~ao Paulo, 
S\~ao Paulo-SP,  05314-970, Brazil
}

\begin{abstract}

We study the effects of distinct types of quenched disorder in the 
contact process (CP) with a competitive dynamics on bipartite 
sublattices. In the model,  the particle creation depends on its first and second
neighbors and the extinction increases according to the local density. 
The clean (without disorder) model exhibits
three phases: inactive (absorbing), active symmetric and active asymmetric,
where the latter exhibits distinct sublattice densities. 
These phases are separated by continuous
transitions; the phase diagram
is reentrant. By performing mean field analysis  
and Monte Carlo simulations we show that symmetric 
disorder destroys the sublattice ordering and therefore
the active asymmetric phase is not present. On the other 
hand, for asymmetric disorder (each sublattice presenting
a distinct dilution rate)  the phase transition occurs between the 
absorbing and the active asymmetric phases. The universality class of 
this transition is governed by the less disordered sublattice.

\end{abstract}




\maketitle

\section{Introduction}

In nonequilibrium systems, an absorbing-state phase transition occurs 
when a control parameter such as a creation or annihilation rate is 
varied, and
the system undergoes a phase transition from a fluctuating state to a 
frozen state, with no
fluctuations (the ``absorbing" state). Absorbing-state phase transitions 
have attracted considerable interest in recent years since they are related 
to the description of several phenomena  such as population dynamics, 
epidemic spreading, chemical reactions, and 
others \cite{marro, hinrichsen,odor04,henkel}.  
During the last decade, several experimental realizations,e.g.,
in turbulent liquid crystals~\cite{take07}, driven 
suspensions~\cite{pine},
superconducting vortices~\cite{okuma} and open quantum 
systems \cite{quantum} have highlighted the importance of
this kind of phase transition.

In analogy with equilibrium phase transitions \cite{goldenfeld}, 
it is expected that the critical phase transitions into absorbing states 
belong to a finite number of universality classes \cite{odor04}. 
However, a complete classification of these nonequilibrium classes is 
still lacking.
In general, absorbing-state transitions in models with short-range 
interactions and lacking a conserved quantity
or symmetry beyond translational invariance belong to the directed 
percolation (DP) universality class \cite{gras-jans}.
On the other hand, models presenting
two absorbing states linked by particle-hole symmetry are known fall
in the voter model universality class \cite{voter}. There are also models
that are free of absorbing states but cannot achieve thermal
equilibrium because their transition rates violate the detailed
balance. An example is the majority vote model \cite{majority}. In
its ordered phase a $Z_2$ symmetry is spontaneously broken,
leading to Ising-like behavior for spatial dimensions $d\geq2$.

A few years ago, a spatially structured model that suffers a phase transition to a single
absorbing state and also exhibit a broken-symmetry phase was proposed \cite{cpsl}.
The model is based in a contact process (CP), where, besides the standard particle creation and annihilation
dynamics, one includes a creation between second neighbors sites
and an annihilation proportional to the local density. This results,
in addition to the usual absorbing and active (symmetric) phases, in an 
unusual active asymmetric phase in which
the distinct sublattices are unequally populated.
A spontaneous symmetry breaking characterizes the reentrant phase 
transition between the symmetric and asymmetric phases. 
Mean-field theory (MFT) and simulations revealed the absorbing 
phase transition belongs to
the directed percolation (DP) class, 
whereas the transitions between active phases
fall into the Ising universality class, as expected from symmetry 
considerations. The symmetry-breaking phase transition was proven to be robust
for different sublattice interactions \cite{salete} and  low diffusion particle
rates \cite{cpsld}.

The inclusion of disorder can
affect the critical behavior of nonequilibrium phase 
transitions dramatically. In real systems, quenched disorder is observed in the form of impurities
and defects \cite{hinrichsen1}, whereas in a regular lattice it can be included 
in the forms of random deletion
of sites or bonds  \cite{noest,adr-dic,vojta06} or  random spatial variation
of the control parameter \cite{durrett,salinas08}.
According to the Harris' criterion \cite{harris}, quenched disorder is a relevant
perturbation, from the field-theoretical point of view, if
$d\nu_\perp < 2$, 
where $d$ is the dimensionality and $\nu_\perp$ is the correlation length exponent of the
pure model.
In these cases, 
quenched uncorrelated randomness induces the emergence 
of rare regions, 
typically located $\lambda_c(0) < \lambda < \lambda_c$, the $\lambda_c(0)$
and $\lambda_c$ being the critical point for the pure and disorder models, 
respectively. Although globally the whole system is constrained
in the subcritical phase, local supercritical regions emerge
due to the presence of the disorder. The lifetime of such 
``active rare regions''  grows exponentially with the domain size,
usually leading to a slow dynamics,
characterized for nonuniversal exponents toward the extinction 
for some interval of the control parameter below criticality. 
This behavior characterizes a Griffiths phase (GP), and was verified
in DP models with uncorrelated disorder irrespective to
the disorder strength \cite{oliveira,vojta09}. In addition, it was
shown this behavior corresponds to the universality
class of the random transverse Ising model \cite{oliveira,vojta09}.
However, some kinds of correlated disorder do not alter the critical
behavior \cite{oliveira2,oliveira3,vojtaprl}. 

In this work, we provide a step further  
by investigating the effects of quenched disorder in 
the phase diagram of the contact processes on sublattices. 
Following the Harris'criterium, disorder should be relevant for the absorbing phase transition, since $\nu_\perp$ =
0.734(4), for $d = 2$ in the clean system \cite{oliveira}. On the other hand, for the symmetry breaking phase transition, 
the Harris criterion is inconclusive, because it corresponds to a
marginal case ($\nu = 1$ for the pure Ising model in $d=2$) \cite{puli}.
Here, we study distinct kinds of disorder, (i) a random homogeneous 
 (ii) inhomogeneous deletion of sites, in which the 
disorder strength is different in each sublattice. Interesting, 
we show, through mean-field analysis and Monte Carlo simulations, 
that each one of the above disorder prescriptions yields completely 
different outcomes.  

The remainder of this paper is organized as follows. In the next
section we review the model and analyze its mean-field theory.
In Sec. III we present and discuss our simulation results; Sec. IV is
to summarize our conclusions.

\section{Model and Mean-Field Theory}

Consider a stochastic interacting particle system, defined 
on a square lattice of linear size $L$, where each site can be 
either occupied by a particle
 or empty. Each particle creates a new particle in one of its 
first-neighbor sites with rate $\lambda_1$, and in one of its 
second-neighbor sites with rate $\lambda_2$. 
Note that in such {\it bipartite} sublattice, $\lambda_1$ is 
the creation rate in the opposite sublattice, whereas $\lambda_2$ 
is the rate in the same sublattice as the replicating particle. Therefore, 
{\it unequal} sublattice occupancies are favored if $\lambda_2 > \lambda_1$. 
An occupied site is emptied at a rate of unity (independent of its 
neighboring sites). 
In addition to the intrinsic annihilation rate of unity,
an ``inhibition term" proportional to the local
density is included in the dynamics. As a consequence of this 
term, if the occupation fraction $\rho_A$ of sublattice $A$ is 
much larger than that of sublattice $B$, for instance, then any
particles created in 
sublattice $B$ will die out quickly, stabilizing the unequal
sublattice occupancies. 

In order to typify the model properties,
we evaluate the macroscopic particle densities of each sublattice 
$A$ and $B$ given by  $\rho_A$ and $\rho_B$, respectively.
In the absorbing (AB) and active symmetric (AS) phases,
$\rho_A=\rho_B=0$ and $\rho_A=\rho_B \neq 0$, respectively. Hence,
$\rho=\rho_A+\rho_B$ is a reliable order parameter for
 absorbing phase transitions. Conversely,
for the active asymmetric (AA) phase, 
it is convenient to calculate the difference of sublattice occupation
by $\phi=|\rho_A-\rho_B|$, since $\phi$ distinguishes 
from the AS phase, where $\phi=0$.

The disorder is introduced by means of a fraction $\Gamma$
of random deletion of sites.
We shall consider two cases: the symmetric and asymmetric, in which the
sublattice remotion is equal ($\Gamma_A=\Gamma_B$) and
different ($\Gamma_A\neq\Gamma_B$), respectively.
In both cases, the disorder is quenched in space and time, i.e, its position or strength
does not change during the evolution of the process.
In order to achieve a qualitative portrait of the phase diagram, we begin by employing
the one-site mean-field theory (MFT). For a lattice with coordination number $q$, 
it results in following coupled equations  

\begin{equation}
\frac{d \rho_A}{dt} = - \left[1 + \mu q^2 \rho_B^2\right] \rho_A + (\lambda_1 \rho_B + \lambda_2 \rho_A)
(1-\Gamma_A-\rho_A)
\end{equation}
and
\begin{equation}
\frac{d \rho_B}{dt} = - \left[1 + \mu q^2 \rho_A^2\right] \rho_B + 
(\lambda_1 \rho_A + \lambda_2 \rho_B)
(1-\Gamma_B-\rho_B).
\end{equation}

\begin{center} 
\begin{figure}[h]
\includegraphics[clip,angle=0,width=1.0\hsize]{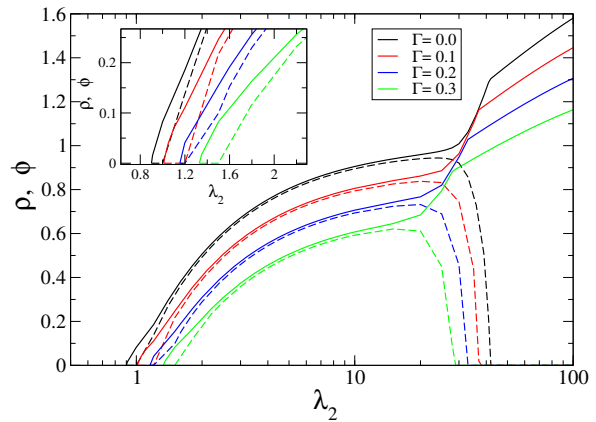}
\caption{{\footnotesize (Color online) Mean-field 
densities $\phi$ (solid curves) e $\rho$ (dahsed) 
for $\lambda_1=0.1$ and $\mu=2.0$. Inset: detail of the data close 
to the absorbing transition (linear scale).}}
\end{figure}
\end{center}

Let us first consider that the disorder is homogeneously
distributed in both sublattices, i.e., $\Gamma_A=\Gamma_B=\Gamma$. 
In this case, one  derives explicit solutions for the densities as
\begin{equation}
\rho=\frac{1}{2k}\left[\sqrt{\left(\frac{\lambda_1+\lambda_2}{2}\right)^2+4k\left[(\lambda_1+\lambda_2)(1-\Gamma)-1\right]}-\frac{\lambda_1+\lambda_2}{2}\right] 
\end{equation}
and
\begin{equation}
\phi=\sqrt\frac{-\left[(\lambda_2-\lambda_1)(1-\Gamma)-1-\lambda_2\rho-k\rho^2\right]}{k} .
\end{equation}

The mean-field densities $\rho$ and $\phi$, are plotted
as function of $\lambda_2$, for distinct values of disorder and 
fixed $\lambda_1=0.1$ in Fig.1. For all fractions of disorder, the system undergoes a phase 
transition to the absorbing state at
the threshold $\lambda_2=\lambda_{2,c}^{ABS}(\Gamma)$, with $\rho=0$,
to the active symmetric phase, where $\rho>0$ and $\phi=0$. Note the
phase transition moves for larger values
of $\lambda_2$  when increasing the disorder, as expected. A further 
increase of $\lambda_2$ gives rise to the AS-AA and AA-AS phase transitions 
at $\lambda_2=\lambda_{2,c}^I(\Gamma)$ and
 $\lambda_2=\lambda_{2,c}^{II}(\Gamma)$, respectively.

\begin{figure}[h]
\includegraphics[clip,angle=0,width=1.0\hsize]{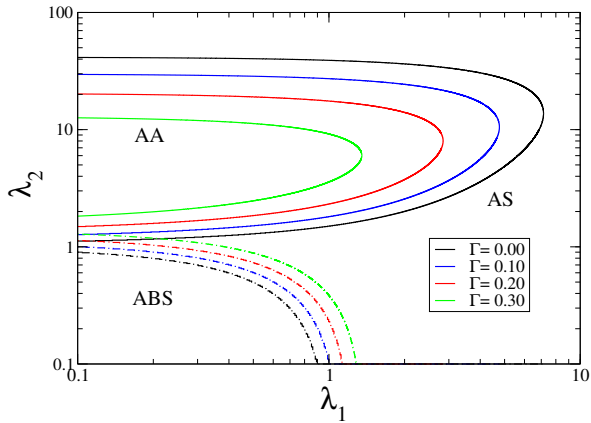}
\caption{{\footnotesize Phase diagram in the $\lambda_1-\lambda_2$ plane for $\mu = 2$, showing absorbing (ABS), 
active-symmetric (AS) and active asymmetric (AA) phase, for distinct values of disorder $\Gamma$.}}
\end{figure}

The phase diagram for distinct values of $\Gamma$ is shown in
Fig. 2.  The increase of $\Gamma$ enlarges the absorbing phase and then
the AB-AS phase transition moves for larger values  of $\lambda_{2,c}^{ABS}$. In
the active phase, the phase diagram is reentrant, and we note
the size of the AA phase is reduced when the disorder increases. 
\begin{figure}[h]
\includegraphics[clip,angle=0,width=1.0\hsize]{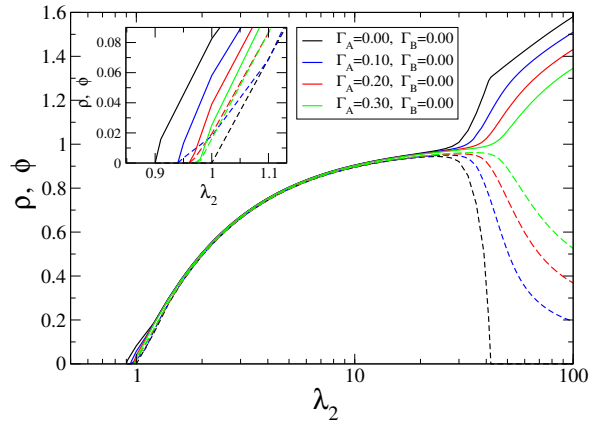}
\caption{{\footnotesize Mean-field densities $\phi$ (solid curves) 
e $\rho$ (dahsed) for $\lambda_1=0.1$ and $\mu=2.0$ with asymmetric 
disorder, $\Gamma_A\neq\Gamma_B$. Inset: detail of the data close 
to the absorbing transition (linear scale).}}
\end{figure}

The asymmetric disorder case, $\Gamma_A\neq\Gamma_B$, is shown in Fig. 3.
MFT indicates the suppression of the AS phase, signed by a smooth change
of both of $\rho$ and $\phi$ at the same $\lambda_{2,c}$. 
The AS phase is not stable for any value of $\lambda_2$,
so that $\phi$ does not vanish for finite values of $\lambda_2$.
So, the phase diagram only presents two phases: the absorbing phase
and the active symmetric phase separated by   a continuous phase transition. 

In the next section, we compare the results from MFT with those evaluated
from numerical simulations.

\section{Results and Discussion}

\subsection{Methods}

We performed extensive Monte Carlo simulations of the model on square
lattices with periodic boundaries.
The simulation scheme is as follows. First, a
site is selected at random. If the site is occupied, it creates a particle at one of its
first neighbors with a probability
$p_1=\lambda_1/W$ or at one of its {\em second} neighbors with a probability $p_2=\lambda_2/W$. Here, $W=(1+\lambda_1+\lambda_2+\mu n_1^2)$ is the sum of
the rates of all possible events. With a complementary probability $1-(p_1+p_2)$, the site is vacated.
To improve the efficiency, we choose the sites from a
list containing the currently
$N_{occ}$ occupied sites; accordingly, the time is
incremented by $\Delta t= 1/N_{occ}$ after each event.
  For simulations in the subcritical and critical absorbing regime,
  we sample the quasi-stationary (QS) regime using the simulation
  method detailed in \cite{qssim}.
In order to draw a comparison with the results from MFT, in all cases we take $\mu=2.0$.

\subsection{Symmetric disorder}

We begin by analysing the symmetric case, where $\Gamma_A=\Gamma_B=\Gamma$. In Fig. 4, we show typical configurations observed on the
bipartite lattice for $\lambda_1=0.1$ and  $\lambda_2=0.1$, for
clean $\Gamma=0$ and disordered system $\Gamma=0.2$. 

\begin{figure}[hbt]
\includegraphics[clip,angle=0,width=1.0\hsize]{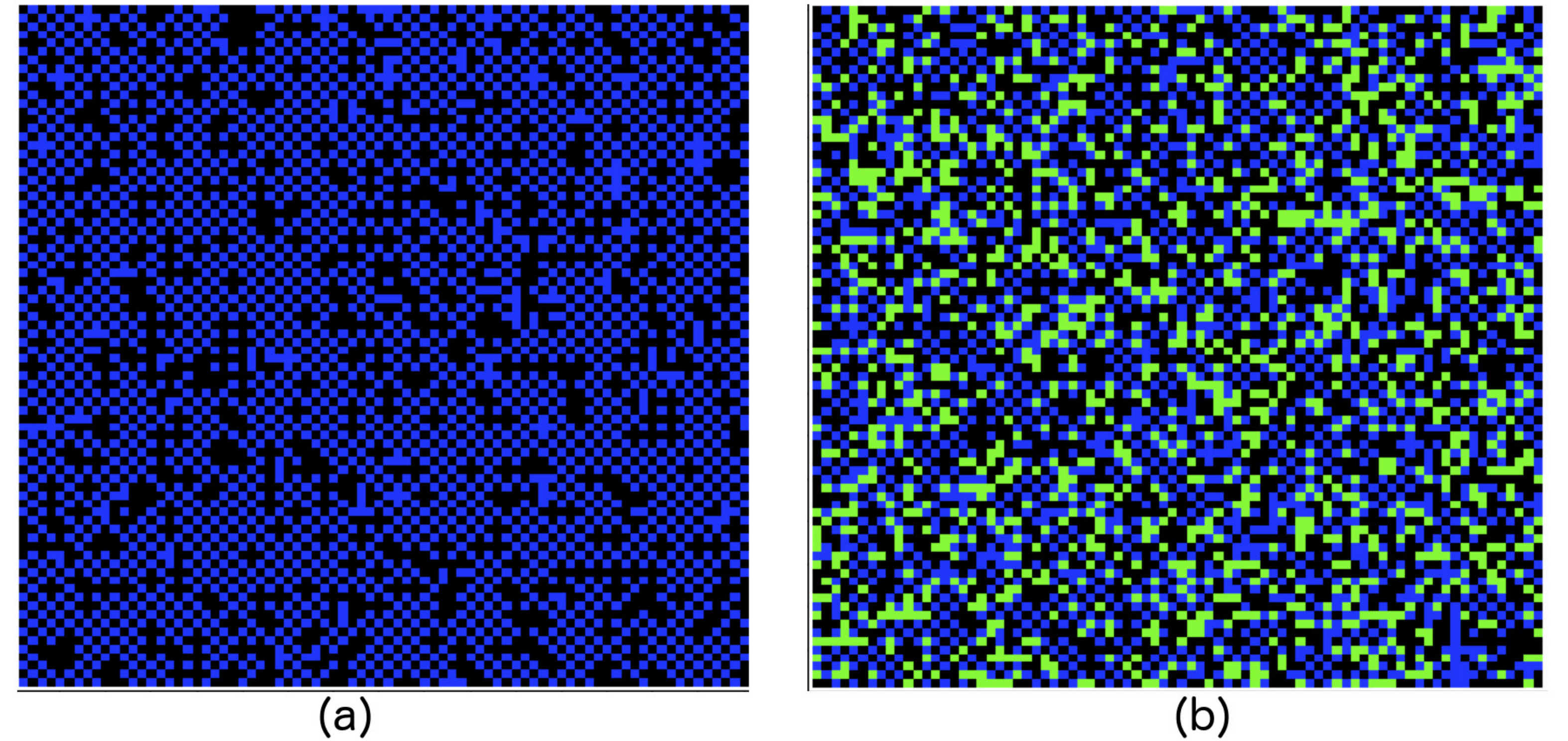}
\caption{{\footnotesize (Color online) Typical configurations observed on the bipartite lattice for $\lambda_1=0.1$ and  $\lambda_2=0.1$, for clean $\Gamma=0.0$ and disordered system $\Gamma=0.2$. Green: inerte sites; Blue: occupied sites and Black: empty sites. Linear system size $L=80$ and $\mu=2.0$.}}
\label{fig2D}
\end{figure}

\begin{figure}[hbt]
\includegraphics[clip,angle=0,width=1.0\hsize]{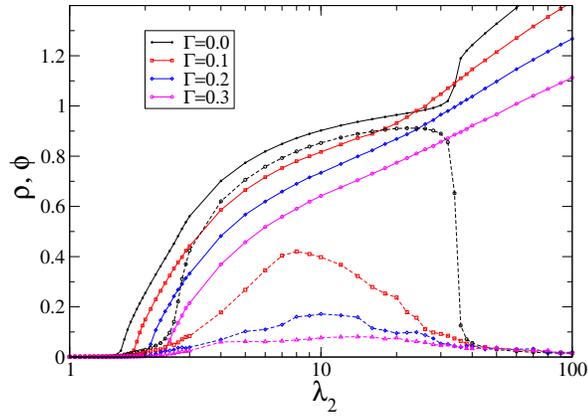}
\caption{{\footnotesize Order-parameters $\rho$ and $\phi$ for distinct (symmetric) $\Gamma$'s vs. $\lambda_2$
, for $\mu = 2$ and $\lambda_1=0.1$. Linear system size $L=80$.}}
\end{figure}

\begin{figure}[h]
\includegraphics[clip,angle=0,width=1.05\hsize]{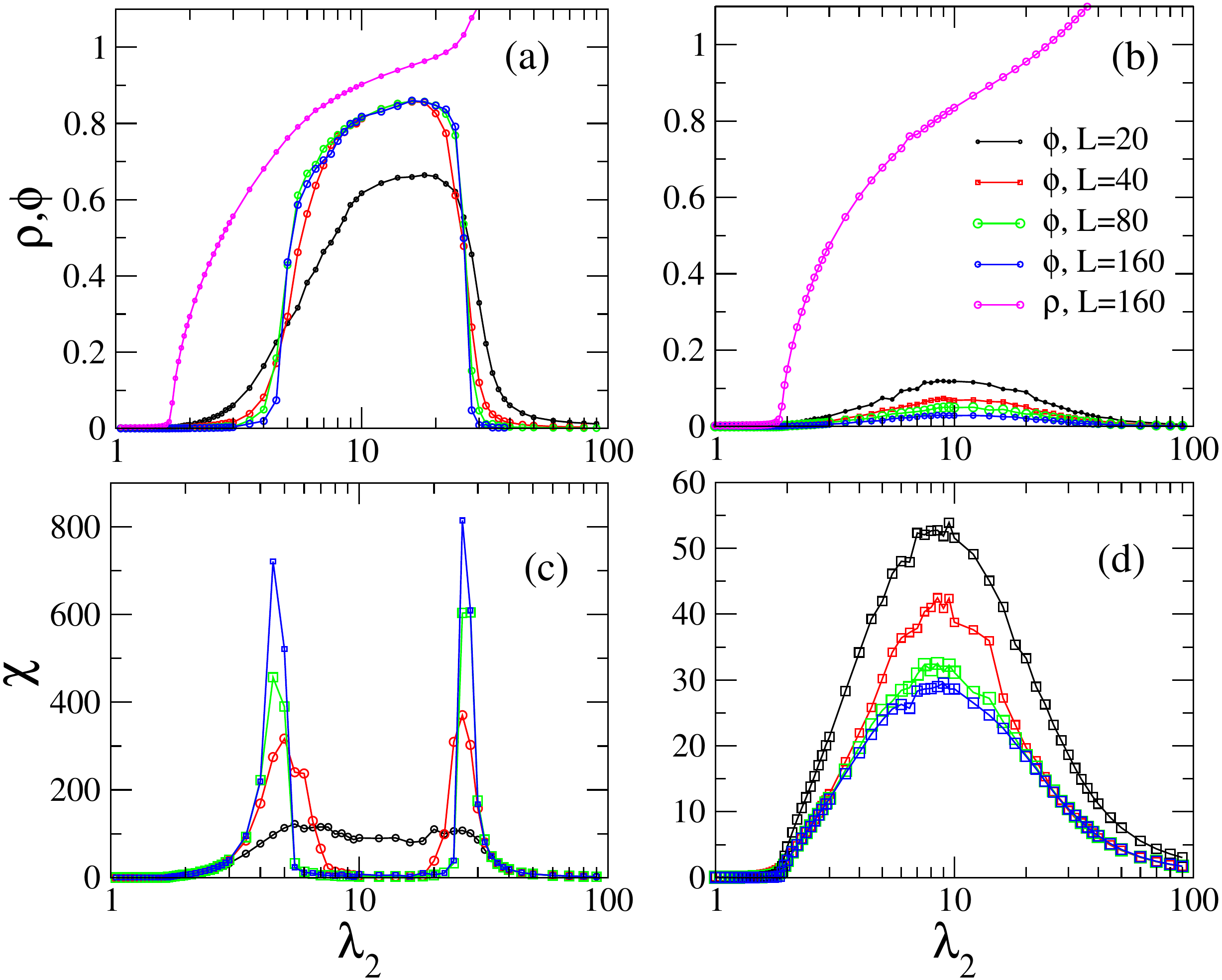}
\caption{{\footnotesize  Quasi-stationary densities $\phi$ and $\rho$ vs. $\lambda_2$,
for $\mu = 2$ and $\lambda_1=0.1$, for the clean system (a), and for $\Gamma=0.1$ (b). Scaled variance $\chi$ of the order parameter
$\phi$  vs. $\lambda_2$for $\mu = 2$ and $\lambda_1=0.1$, for the clean system (c), and for $\Gamma=0.1$ (d). }}
\end{figure}

\begin{figure}[h]
\includegraphics[clip,angle=0,width=1.0\hsize]{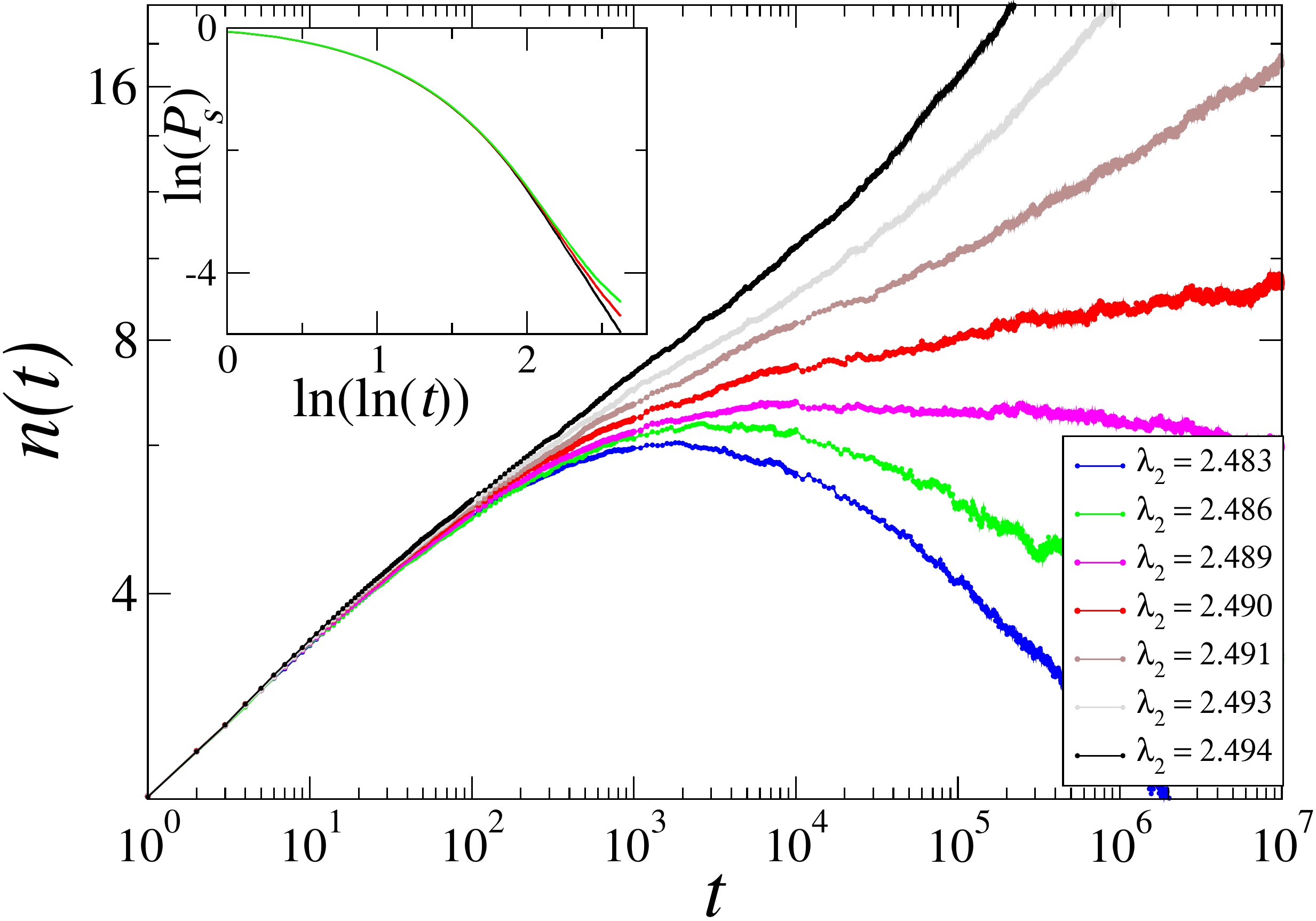}
\caption{{\footnotesize Number of active particles $n$
for $\Gamma=0.3$, $\mu = 2$ and $\lambda_1=0.1$. Inset: survival probability for $\lambda_2=2.489$, $2.490$ and 2.491.}}
\label{fig}
\end{figure}

\begin{figure}[h]
\includegraphics[clip,angle=0,width=1.0\hsize]{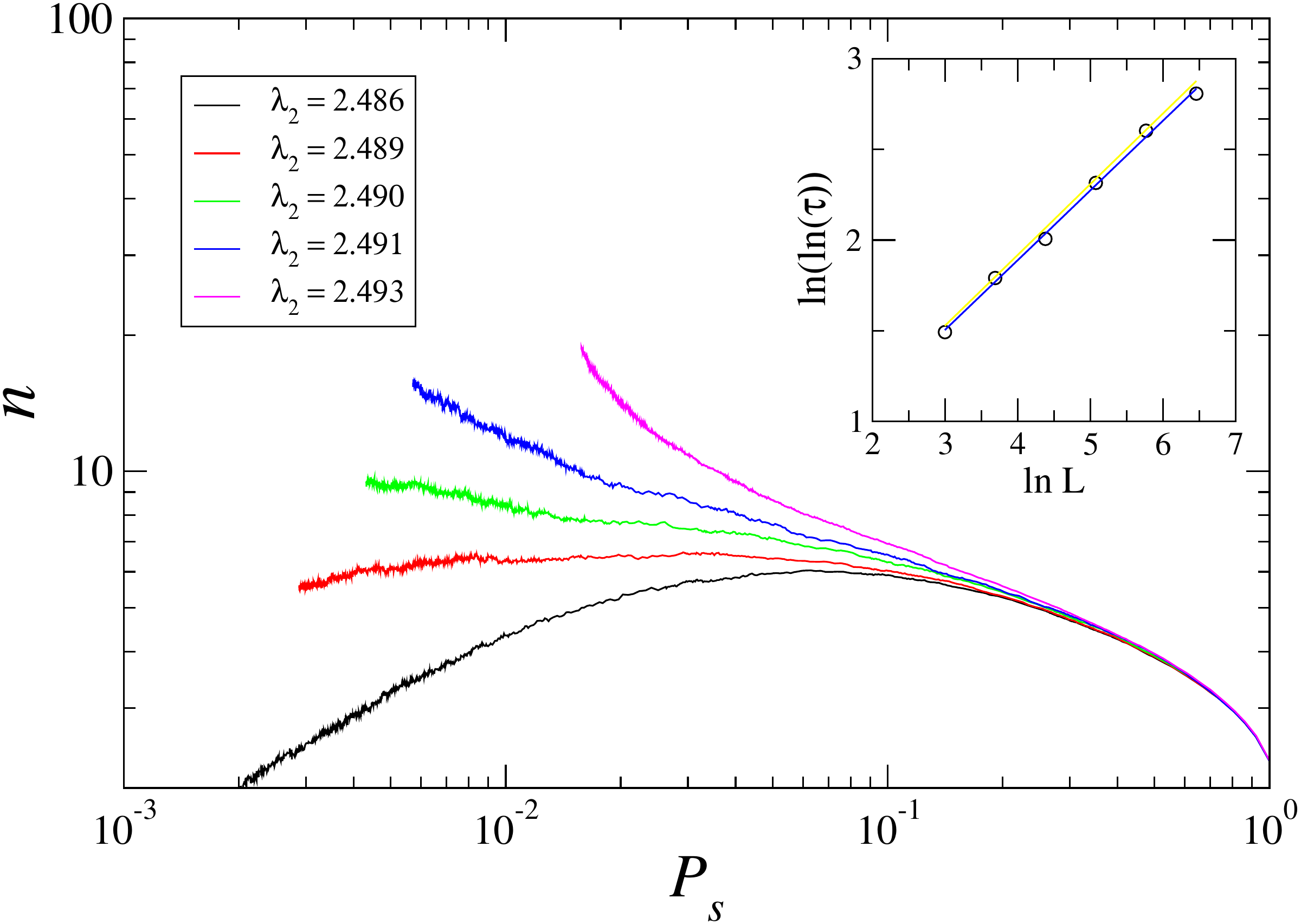}
\caption{\footnotesize{. Log-log plot of the number of ocuppied sites $n$ as a function of the survival
probability $P_s$ obtained from spreading simulations. Inset: Finite-size scaling of the lifetime of the QS state $\tau$.
for $\Gamma=0.3$, $\mu = 2$ and $\lambda_1=0.1$. }}
\label{fig}
\end{figure}

\begin{figure}[h]
\includegraphics[clip,angle=0,width=1.0\hsize]{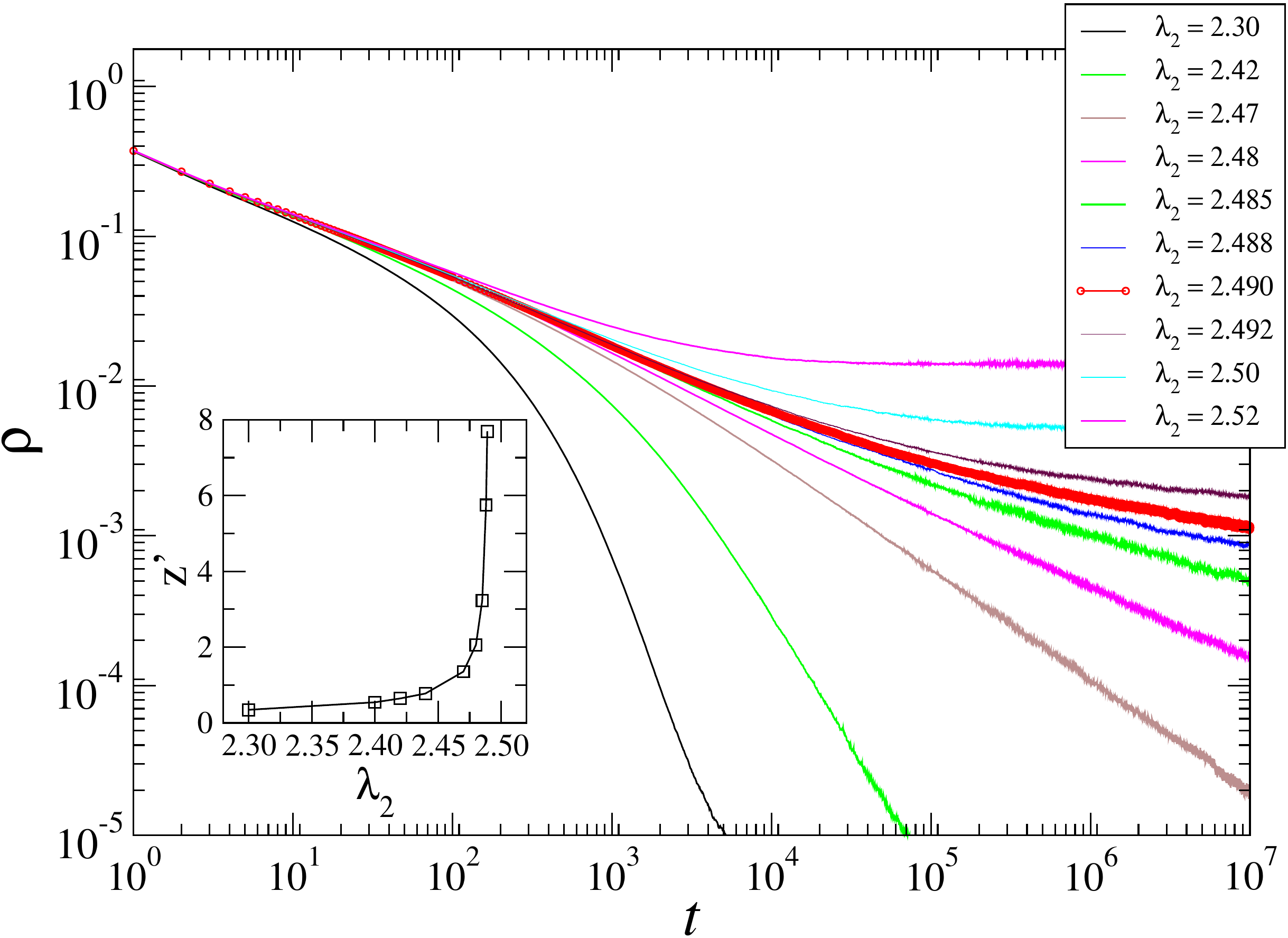}
\caption{\footnotesize{Initial decay simulations: particle density $\rho$ vs. $t$ for $\mu = 2$ and $\lambda_1=0.1$ 
for symmetric disorder case, with $\Gamma=0.3$ (linear system size $L=2000$). Inset: Dynamical
exponent $z'$ as a function of $\lambda_2$.}}
\label{fig}
\end{figure}

Figure 5 shows the densities of $\rho$ and $\phi$ on a square lattice
with linear system size $L=80$. As expected, the absorbing
phase transition occurs at a higher value of $\lambda_2$ when
disorder is introduced. We also observe that increasing the
fraction of disorder reduces the possibility of an AA phase, since
the values of $\phi$ reduces when we increase the fraction of disorder.  

In order to clarify the effects of disorder on the stability of the
AA phase, we analyze the dependence of the order parameter $\phi$
with the linear system size $L$.
In Fig. 6, we compare the clean model with the disordered system,
with $\Gamma=0.1$.  Panels (a) and (b) show the order parameter
$\rho$ and $\phi$ versus
 $\lambda_2$ for distinct systems sizes. While for the clean version the
 the AA phase and the reentrant phase transition are
observed for all system sizes (see panel (a)),
the disordered system shows a remarkably different
picture, in which the sublattice occupations
become equivalent for large  $L$.
Therefore, we conclude that $\phi$ vanishes when $L\to \infty$ and the AA phase is suppressed
by the disorder.
Such behavior is reinforced by  analyzing the order parameter
variance $\chi=L^2(\langle\phi^2\rangle-\langle\phi\rangle^2)$, as
shown in panels $(c)$ (clean) and $(d)$ (disordered). In contrast
to the  clean system, in which $\chi$ diverges nearby the 
transitions AB-AA and AA-AB as $L \to \infty$, no divergence
is verified in such case.
So, in contrast to the MFT predictions, we observe that symmetric disorder
forbids the stability of AA phase and 
therefore the disordered system does not show symmetry breaking. 

Now, let us characterize the absorbing phase transition in the
disordered system. For locating the
critical creation rate $\lambda_{2,c}$, 
 we study the time evolution of the number of active particles $n(t)$
 starting from an initial configuration the simulation with one pair of
 neighboring active particles. The
 critical value $\lambda_{2,c}$
 can be estimated as the threshold $\lambda_2$ separating
 asymptotic growth from the decay towards the
 absorbing phase. 
 
 Figure 7 exemplifies  the results
 for $\Gamma=0.3$ in which $\lambda_{2,c}=2.490(1)$.  
We observe that, analogously to the diluted CP, the critical behavior
presents activated dynamic scaling, with 
\begin{equation}  
n(t)\sim \ln(t/t_0)^\theta
\end{equation}
and with the survival probability 
\begin{equation} 
P_s\sim \ln(t/t_0)^{-\delta}.
\end{equation}
From the data in Fig.7, we estimate $\ln(t_0)=3.0(5)$, and $\theta=0.21(4)$, in agreement 
with the value $\theta=0.15(3)$, obtained in \cite{vojta09}.
From the data shown in the inset of  Fig. 7, we find $\delta=2.1(3)$,
very close to the value $\delta=1.9(2)$ observed for the usual CP with
random dilution, indicating that critical behavior
of the disordered model  belongs to the universality class
of the diluted CP \cite{vojta09,hada}.

Combining the activated scaling relations for $n(t)$ and $P_s(t)$, eqs. (5) and (6), we find that 
\begin{equation}
n\sim P_s^{-\delta/\theta}
\end{equation}
 at criticality. This relation does not depend on $t_0$,
and is useful to check our estimate of the critical point. 
In Fig.8, we plot $n$ as function of $P_s$, and observe a power law that at
$\lambda_2=2.490$, while the curves for $\lambda_2=2.491$ ($\lambda_2=2.489$) 
veers up (down), thus confirming the accuracy of our estimate for the critical point. At criticality, 
we found the exponent ratio $\delta/\theta=0.09(2)$, in coherence with with the values of 
$\delta$ and $\theta$ obtained from the data in Fig.7. This value is also in agreement with the value
$\delta/\theta=0.075(5)$ obtained for the diluted contact process in \cite{hada}.

In the active-dynamical scenario, the lifetime of the process
follows 
\begin{equation}
\ln \tau \sim L^{\psi},
\end{equation}
 where $\psi$ is
a universal exponent. From the data in the inset of Fig 8, we
found $\psi=0.44(5)$, close to the value $\psi= 0.51(6)$
obtained for the diluted CP \cite{vojta09}.

Finally, another important effect that disorder can induce in absorbing
phase transitions is the emergence of Griffiths phases. A Griffiths phase is a region inside the subcritical phase where 
the long-time decay of $\rho$ towards the extinction 
is algebraic (with non-universal exponents), rather than exponential. In Fig.9,
we present results from initial decay simulations, where the system starts its dynamics
from a fully occupied lattice. We observe the existence of a range of values of $\lambda_2$ in the subcritical regime 
 where the long-time
behavior of the density decays as a power law, 
\begin{equation}
\rho(t)\sim t^{-2/z'}
\end{equation}
with the non-universal dynamical exponents $z'$ following 
\begin{equation}
z'\sim|\lambda_2-\lambda_{2,c}| ^{-\psi\nu_\perp},
\end{equation}
where $\lambda_{2,c}$ is the critical value of the control parameter $\lambda_2$ \cite{vojta09}.
In the inset of Fig. 9 we observe that $z'$ diverges when
$\lambda_2$ approaches $\lambda_{2,c}= 2.490$. From these data, 
we obtain $\psi\nu_\perp=0.60(1)$, in good agreement with the value
$\psi\nu_\perp \approx 0.61$ reported in \cite{vojta09} for the diluted CP.

 In resume, our results for the effects of symmetric disorder are in agreement with the predictions from the Harris criterium, since $\nu_\perp$ =
0.734(4), for $d = 2$ in the clean system, and therefore the disorder is relevant for the absorbing phase transition \cite{oliveira}.
Similar trends have been observed for lower values
of $\Gamma$, although larger crossover times toward
the infinite-random behavior are expected in such cases
\cite{oliveira,vojta09}.  

\subsection{Asymmetric disorder}

Now, we investigate the effects of asymmetric disorder in which
disorder strengths is different in each sublattice.
This is exemplified in  Fig. 10 for  $\Gamma_1=0$ and
distinct values of $\Gamma_2$. We note that even 
a very small asymmetric disorder ($\Gamma_2=0.05$),
the AS phase is suppressed and only less disordered sublattice is populated. 
Therefore, only a phase transition  between the
inactive and the asymmetric phase is presented, whose 
critical behavior is ruled by the less disordered sublattice. 
In all these cases, we observe that at criticality, $\rho\sim L^{-\beta/\nu_\perp}$ and the lifetime $\tau\sim L^{-z}$. For example, for $\Gamma_2=0.2$,
shown in Fig 11,  
we find $\beta/\nu_\perp= 0.81(2) $ and $z=1.75(3) $, in agreement
with the DP values $\beta/\nu_\perp=0.797(3) $ and $z=1.7674(6)$. The
inset shows the moment ratios of the order parameter, 
$m=\langle\rho\rangle^2/\langle\rho^2\rangle$ goes to a universal
value $m=1.33(1)$ at criticality, in comparison to the known DP value
$m=1.3264(5)$ \cite{cpsl}.
\begin{figure}[h]
\includegraphics[clip,angle=0,width=1.0\hsize]{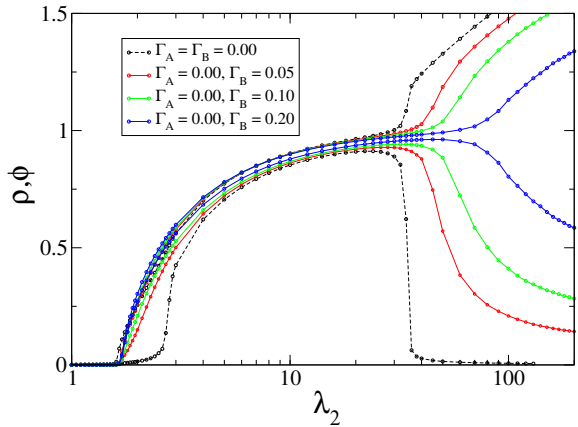}
\caption{\footnotesize{For $L=80$ and assymmetric disorder:
    Order-parameters $\phi$
    and $\rho$ for $\mu = 2$ and $\lambda_1=0.1$
    for distinct $\Gamma_A$ and $\Gamma_B$.}}
\label{fig}
\end{figure}
\begin{figure}[h]
\includegraphics[clip,angle=0,width=1.1\hsize]{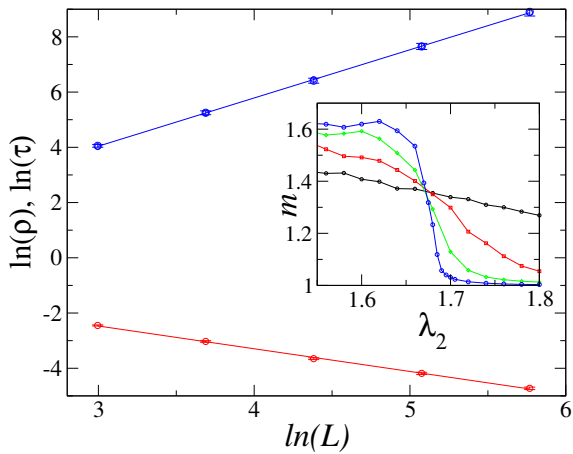}
\caption{\footnotesize{ Assymetric disorder, with $\Gamma_1=0.0$ and $\Gamma_2=0.2$:  Scaled critical QS density of active sites $\ln\rho$ (bottom)
and scaled lifetime of the QS state $\ln\tau$ (top), versus $\ln L$. Inset: Moment ratio $m$ versus $\lambda_2$.  on a square lattice.
Parameters: $\mu = 2$ and $\lambda_1=0.1$}}
\label{fig}
\end{figure}

\subsection{Phase diagrams}

Our simulation results for the effects of disorder are resumed in the phase diagrams shown in Fig.12. In (a), the clean system exhibits, besides the
absorbing phase, a reentrant active asymmetric phase (with $\phi>0$) inside the active symmetric phase. In \cite{cpsl}, it was shown that 
the active-absorbing phase belongs to the DP universality class, while the AA-AS symmetry-breaking phase transition is Ising-like. 

The effects of asymmetric disorder in the phase diagram  are shown in Fig.12 (b) and (c). We note that the AS phase vanishes, and there are
only two phases, the absorbing and the AA phase. The universality class of the transition between these phases is governed by the less 
disordered sublattice. So, while in the case (b), the phase transition belongs to the DP universality class, in the case (c) it belongs to the
universality class of the diluted CP.

Finally, if the disorder is symmetric, as in case (d), we note that the AA phase vanishes, and the absorbing-AS phase transition 
belongs to the class of the diluted CP. This system exhibits activated scaling and Griffiths phases as shown in detail in section III.B. 

\begin{figure}[h]
\includegraphics[clip,angle=0,width=1\hsize]{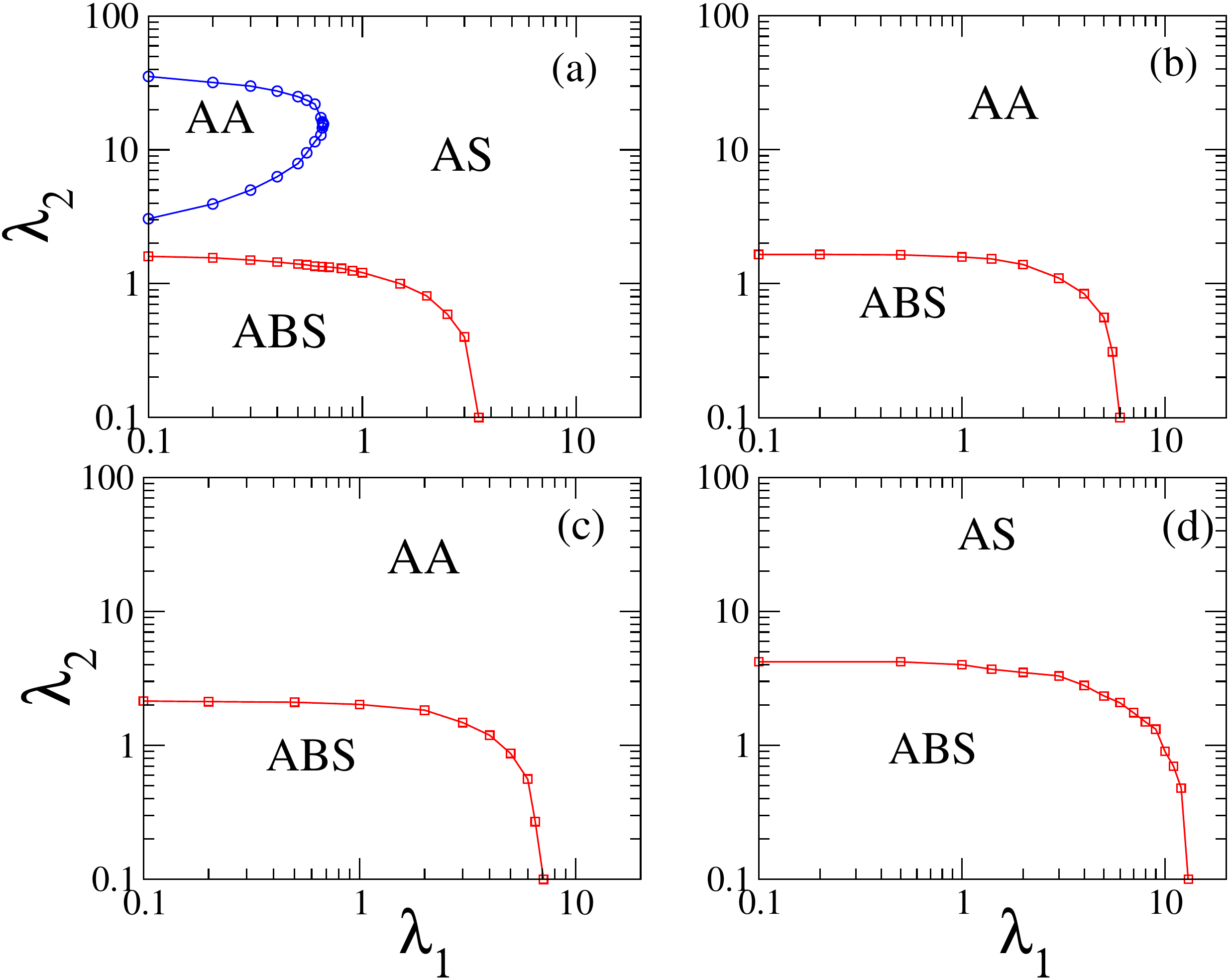}
\caption{{\footnotesize Phase diagrams for (a) clean system, (b) $\Gamma_A=0 $ and $\Gamma_B=0.3$, 
(c) $\Gamma_A=0.1 $ and $\Gamma_B=0.3$, and (d) $\Gamma_A= \Gamma_B=0.3$ ( $\mu = 2$ in all cases).}}
\end{figure}

\section{Conclusions}

In this work, we have investigated the effects of quenched disorder
in the phase diagram of the competitive contact processes on sublattices.
Through mean-field analysis and Monte Carlo simulations, we have studied distinct
types of disorder, (i) a random {\it homogeneous} deletion of sites (ii) an
asymmetrical disorder, in which the disorder strength is different in
each sublattice. Interesting, each one of the above disorder
prescriptions yields completely different outcomes. 

We observe that in the case (i), the disorder destroys the
asymmetric active phase, and therefore, the symmetry-breaking
phase transition. So, there are only two phases, the absorbing and the active (symmetric) phases. 
The absorbing-active phase
transition exhibited belongs to the universality class of
the disordered contact process. Effects related to an infinite randomness fixed point
are observed, such as activated dynamics and Griffiths phases in the subcritical regime.

A distinct behavior is observed if each sublattice has a different disorder strength. 
In such case, the symmetrical active phase is not stable, and we observe a phase
transition directly from the active asymmetric phase to the
absorbing phase. The critical behavior in this case is
governed by the sublattice with less disorder, for instance, if one of
the sublattices has no disorder, then the phase
transition will fall in the DP class.

A natural extension of the present work would be the study of the effects of temporal disorder \cite{temp0,temp1,temp2,temp3} on the robustness of the
symmetry-breaking phase transition, and in its the critical behavior.
It is important to mention that, besides the theoretical interest in the field of nonequilibrium phase transitions, suppression
of activity at the nearest neighbors of active sites resembles
biological lateral inhibition, known to be important in the
visual system of many animals \cite{brain}.  Also, our work can be useful to understand the effects of
heterogeneities in extended systems showing {\em checkerboard} pattern distributions such as mutually
 exclusive species co-occurrences \cite{eco}.

\vspace{1cm}

\noindent{\bf Acknowledgments}

This work was supported by CNPq and FAPEMIG, Brazil.

\bibliographystyle{apsrev}

\begin{thebibliography}{100}

\bibitem{marro}
        J. Marro and R. Dickman,
        {\it Nonequilibrium Phase Transitions in Lattice Models}
        (Cambridge University Press, Cambridge, 1999).

\bibitem{hinrichsen}
        H. Hinrichsen,
        Adv. Phys. {\bf 49}, 815 (2000).


\bibitem{henkel}
        M. Henkel, H. Hinrichsen and S. Lubeck,
        {\it Non-Equilibrium Phase Transitions Volume I: Absorbing Phase Transitions}
        (Springer-Verlag, The Netherlands, 2008).

\bibitem{odor04}
        G. \'Odor,
        Rev. Mod. Phys {\bf 76},  663 (2004).

\bibitem{take07}
        K. A. Takeuchi, M. Kuroda, H. Chat\'e, and M. Sano,
        Phys. Rev. Lett. {\bf 99}, 234503 (2007).

\bibitem{pine}
        L. Cort\'e, P. M. Chaikin, J. P. Gollub, and D. J. Pine,
        Nature Physics {\bf 4}, 420 (2008).
        
\bibitem{okuma} S. Okuma, Y. Tsugawa, and A. Motohashi,
        Phys. Rev. B{\bf 83}, 012503 (2011).

\bibitem{quantum} R. Guti\'errez, C. Simonelli, M. Archimi, F. Castellucci, E. Arimondo, D. Ciampini, M. Marcuzzi, I. Lesanovsky, and O. Morsch
Phys. Rev. A {\bf 96}, 041602(R) (2017).

\bibitem{goldenfeld} N. Goldenfeld, {\it Lectures on phase transitions and the renormalization group} (Addison-Wesley, 1992).

\bibitem{gras-jans}
H. K. Janssen, Z. Phys. B {\bf 42}, 151 (1981); \\
P. Grassberger,  Z. Phys. B {\bf 47}, 365 (1982).

\bibitem{voter}
        I. Dornic, H. Chat\'e, J. Chave and H. Hinrichsen,
        Phys. Rev. Lett. \textbf{87}, 045701 (2001).
        
\bibitem{majority}  M. J. de Oliveira, J. Stat. Phys.{\bf  66}, 273 (1992).

\bibitem{cpsl} M. M. de Oliveira and R. Dickman, Phys. Rev. E {\bf 84}, 011125 (2011).

\bibitem{salete} S. Pianegonda and C. E. Fiore, J.  Stat. Mech. 2014,  P05008 (2014).

\bibitem{cpsld} M. M. de Oliveira and C. E. Fiore,  J.  Stat. Mech. 2017,  P053211 (2017).

\bibitem{hinrichsen1} H. Hinrichsen, Braz. J. Phys. {\bf 30}, 69 (2000).

\bibitem{noest}
A. J. Noest, Phys. Rev. Lett. {\bf 57}, 90 (1986).

\bibitem{adr-dic}
Moreira A G and Dickman R, Phys. Rev. E \textbf{54} R3090 (1996); \\
Dickman R and Moreira A G, Phys. Rev. E \textbf{57} 1263 (1998).

\bibitem{vojta06}
T. Vojta and M. Y. Lee, Phys. Rev. Lett. {\bf 96}, 035701 (2006).

\bibitem{durrett} M. Bramson, R Durrett, and R. Schonmann, Ann. Prob. {\bf 19},
960 (1991).

\bibitem{salinas08} M. S. Faria, D. J. Ribeiro and S. A. Salinas, J.
Stat. Mech, P01022 (2008).

\bibitem{harris} A. B. Harris, J. Phys. C {\bf 7}, 1671 (1974).

\bibitem{oliveira} M. M. de Oliveira  and S. C. Ferreira, J. Stat. Mech P11001 (2008).

\bibitem{vojta09} 
 T. Vojta, A. Farquhar and M. Mast, Phys. Rev. E {\bf 79}, 011111 (2009).
 
\bibitem{oliveira2}
     M. M. de Oliveira, S.G. Alves, S.C. Ferreira and R. Dickman,
     Phys. Rev. E {\bf 78}, 031133 (2008).
     
\bibitem{vojtaprl} H. Barghathi and T. Vojta
Phys. Rev. Lett. {\bf 113}, 120602 (2014).     

\bibitem{oliveira3}
     M. M. de Oliveira, S.G. Alves and S.C. Ferreira,
     Phys. Rev. E {\bf 93}, 012110 (2016).
     
\bibitem{puli} P.H.L. Martins and J.A. Plascak, Phys. Rev. E {\bf 76}, 012102 (2007).

\bibitem{qssim}
        M. M. de Oliveira and R. Dickman,
        Phys. Rev. E {\bf 71}, 016129 (2005);
        M. M. de Oliveira and R. Dickman, Braz. J. Phys. 36, 685 (2006).

\bibitem{hada} A. O Hada and M. J. de Oliveira, J.  Stat. Mech.  {\bf 2017}, 043209 (2017).

\bibitem{temp0}  F. Vazquez, J. A. Bonachela, C. L\'opez and M. A. Mu\~noz,
Phys. Rev. Lett. {\bf 106}, 235702 (2011).

\bibitem{temp1}  R. Mart\'inez-Garc\'ia, F. Vazquez, C. L\'opez, and M. A.
Mu\~noz, Phys. Rev. E {\bf 85}, 051125 (2012).

\bibitem{temp2} M.M. de Oliveira and C.E. Fiore,
Phys. Rev. E {\bf 94}, 052138 (2016).

\bibitem{temp3} C.E Fiore, M.M. de Oliveira and J.A. Hoyos,
Phys. Rev. E {\bf 98}, 032129 (2018).

\bibitem{brain} W. Lytton, From Computer to Brain (Springer-Verlag,
New York, 2002).

\bibitem{eco} E.F. Connor, M.D. Collins and D. Simberloff,  Ecology {\bf 94}, 2403 (2013).


\end{thebibliography}

\end{document}